\documentclass[10pt,aps,pre,twocolumn,floatfix,showpacs,superscriptaddress,raggedbottom]{revtex4-1}

\usepackage{ucs}
\usepackage[utf8x]{inputenc}
\usepackage[icelandic,english]{babel}
\usepackage[T1]{fontenc}

\usepackage[caption=false]{subfig}
\usepackage{verbatim}   
\usepackage{color}      

\usepackage{amssymb,amsmath,bm}
\usepackage{graphicx,latexsym}
\usepackage{dcolumn}
\usepackage{ulem}

\newcommand{\ket}[1]{ |#1\rangle}
\newcommand{\braket}[2]{ \langle #1| #2 \rangle}
\newcommand{\brainnf}[3]{ \langle #1 | #2 | #3 \rangle}

\newcommand{\pket}[1]{ |#1)}

\newcommand{\pbrainnf}[3]{ ( #1 | #2 | #3 )}

\newcommand{\expo}[1]{e^{#1}}

\newcommand{\alg}[1]{\left| #1\right|}
\newcommand{\slaufa}[1]{\left\{ #1 \right\}}
\newcommand{\svig}[1]{\left( #1 \right)}
\newcommand{\horn}[1]{\left[ #1 \right]}
\newcommand{\dif}{\,\mathrm{d}}
\newcommand{\ntext}[1]{_{\textrm{#1}}}
\newcommand{\ttext}[1]{^{\text{#1}}}

\begin{document}

\title{Nonperturbative Approach to Circuit Quantum Electrodynamics}

\author{Olafur Jonasson}
\affiliation{Science Institute, University of Iceland,
        Dunhaga 3, IS-107 Reykjavik, Iceland}

\author{Chi-Shung Tang}
\email{cstang@nuu.edu.tw}
 \affiliation{Department of Mechanical Engineering,
  National United University, 1, Lienda, Miaoli 36003, Taiwan}

\author{Hsi-Sheng Goan}
\email{goan@phys.ntu.edu.tw}
 \affiliation{Department of Physics and Center for Theoretical Sciences,
National Taiwan University, Taipei 10617, Taiwan}
 \affiliation{Center for Quantum Science and Engineering,
 National Taiwan University, Taipei 10617, Taiwan}

\author{Andrei Manolescu}
\affiliation{Reykjavik University, School of Science and
Engineering, Menntavegur 1, IS-101 Reykjavik, Iceland}

\author{Vidar Gudmundsson}
\email{vidar@hi.is}
 \affiliation{Science Institute, University of Iceland,
        Dunhaga 3, IS-107 Reykjavik, Iceland}

%

\begin{abstract}
We outline a rigorous method which can be used to solve the many-body
Schrödinger equation for a Coulomb interacting electronic system in an
 external classical magnetic field as well as a quantized electromagnetic
 field. Effects of the geometry of the electronic system as well as the 
polarization of the quantized electromagnetic field are explicitly taken 
into account. We accomplish this by performing repeated truncations of 
many-body spaces in order to keep the size of the many particle basis 
on a manageable level. The electron-electron and electron-photon interactions 
are treated in a nonperturbative manner using ``exact numerical 
diagonalization''. Our results demonstrate that including the diamagnetic 
term in the photon-electron interaction Hamiltonian drastically improves 
numerical convergence. Additionally, convergence with respect to the number 
of photon states in the joint photon-electron Fock space basis is fast. 
However, the convergence with respect to the number of electronic states 
is slow and is the main bottleneck in calculations.
\end{abstract}



\maketitle

%
%

\section{Introduction}

To describe the interaction between matter and a single-mode quantized
electromagnetic field, some version of the Jaynes-Cummings (JC) model
is often applied \cite{jaynes1963}. The JC-model was first employed by
Jaynes and Cummings to describe the interaction of photons with molecules
but since then it has also been used in cavity electrodynamics to 
successfully describe matter-photon interaction in semiconductor 
nanostructures such as quantum dots \cite{Kasprzak2010} and in 
superconducting qubits \cite{Wallraff2004,Fink2008}. Advances in the field
of circuit quantum electrodynamics have enabled us to enter the ultrastrong
coupling regime where the photon-matter coupling strength reaches a 
considerable fraction of the energy of a single cavity photon. This has 
been achieved by taking advantage of large dipole moments and long coherence
times in superconducting flux qubits 
\cite{Devoret2007,Niemczyk2010,Abdumalikov2008} and semiconductor quantum
wells \cite{Gunter2009,Anappara2009,Ciuti2005} embedded in high quality
micro-cavities.

In the ultrastrong regime, the JC model fails and evidence of the breakdown
of the JC-model with the rotating wave approximation has been observed
experimentally in superconducting \cite{Niemczyk2010} and semiconductor
systems \cite{Gunter2009,Anappara2009}. Exact numerical calculations predict
the failure of the JC-model (even without the rotating wave approximation)
where the effects of the diamagnetic matter-photon interaction term as well
as effects of states which are not part of the two level system approximation
come into play with high coupling strength \cite{jonasson2012}.

Using the method described later in this publication, we have been able to
calculate time dependent electron transport through a photon cavity
\cite{gudmundsson2012} and to test the validity of the Jaynes-Cummings
model in the ultrastrong coupling regime \cite{jonasson2012}. With our
approach, it would be relatively easy to add a time dependent perturbation
to the closed system and integrate the equation of motion numerically.
Choosing the frequency of the perturbation such that the EM field does
not have time to adjust adiabatically, it is possible to investigate
non-adiabatic dynamics related to the dynamical Casimir effect
\cite{casimir1948} where photons can then be excited out of vacuum in
correlated pairs. This non-adiabatic effect was recently observed
experimentally for the first time \cite{wilson2011}.

In this paper we describe a general method which can be used to describe
the interaction between an electronic/atomic system with a single-mode
quantized electromagnetic field. We begin by calculating eigenfunctions
and energies of the single-electron Hamiltonian (initially completely ignoring
many-body effects and the EM field). We then use a number of the lowest
single-electron eigenstates to construct a many-electron Fock state basis
which is used to compute the eigenstates and energies of the many-electron
Hamiltonian including the Coulomb interaction between electrons. Finally,
we use a number of the lowest Coulomb interacting eigenstates to construct
a joint electron-photon basis. In diagonalizing the electron-photon
Hamiltonian we obtain its eigenstates which include the electron-photon
and electron-electron interaction ``exactly'' in the sense that the only
approximations are the finite sizes of single/many particle bases and finite
size of grids used for numerical integration. The results are convergent with
respect to these parameters in a controllable manner.

The paper is organized as follows. In Sec. \ref{sec:single} we give a
description of the single-electron Hamiltonian and calculate its
eigenfunctions, which we use as a basis for many-body calculations.
In Sec. \ref{sec:many} we introduce the second quantization many-body
formalism needed to account for the Coulomb interaction between electrons.
In Sec. \ref{sec:EM} we couple the electronic system to single-mode quantized
electromagnetic field and solve the many-body Schrödinger equation using a
basis of Coulomb interacting electron states as well as photon Fock states.
Results and concluding remarks are presented in Secs. \ref{sec:results} and
\ref{sec:conclusions} respectively.

\section{Single-electron Hamiltonian}
\label{sec:single}

The system under investigation is a two-dimensional electronic nanostructure
exposed to a static (classical) external magnetic field at a  low temperature.
The electronic nanostructure is assumed to be fabricated by split-gate
configuration  in the y-direction, forming a parabolic confinement with
the characteristic frequency $\Omega_0$ on top of a semiconductor
heterostructure. The ends of the  nanostructure in the x-direction at
$x = \pm L_x / 2$ are etched, forming a hard-wall confinement of length $L_x$.
The external classical magnetic field is given by $\mathbf B=B\mathbf{\hat{z}}$
with a vector potential $\mathbf A=(-By,0,0)$. Since we are interested in
geometrical effects, we need the single-electron eigenstates to construct
a many-body basis. We therefore need to solve the time independent Schrödinger
equation for the Hamiltonian
\begin{align}
 \nonumber
 H_0 & = \frac{1}{2m}(\mathbf p + q\mathbf A)^2 + \frac12m\Omega_0^2y^2\\ \label{eq:shr}
     & = \frac{1}{2m} p_x^2 + \frac{1}{2m} p_y^2 + \frac12 m\Omega_w^2 y^2 + i\omega_c y p_x\ ,
\end{align}
where $m$ is the effective mass of an electron, $-q$ its charge, $\mathbf p$
the canonical momentum operator, $\omega_c = qB/m$ is the cyclotron frequency
and $\Omega_w = \sqrt{\omega_c^2+\Omega_0^2}$ is the modified parabolic
confinement. Note that the spin degree of freedom is neglected. With the
boundary conditions $\psi(\pm L_x/2,y)=\psi(x,\pm \infty )=0$, the mixing
term $i\omega_c y p_x$ makes it impossible to use separation of variables
to solve the time independent Schrödinger equation for the Hamiltonian in
Eq. \eqref{eq:shr} analytically. This means we will have to resort to
numerical techniques. This procedure is relatively straightforward and
will only be briefly covered here.

To solve the time independent Schrödinger equation for $H_0$, we compute the
matrix representation of $H_0$ in the basis
$\{ \ket{\phi_n} \otimes \ket{\varphi_m} \}$ where
$\ket{\phi_n} \otimes \ket{\varphi_m}$ are eigenstates of $H_0$ when the mixing
term $i\omega_c y p_x$ is omitted. The matrix elements are calculated
analytically. Furthermore let us assume we have a bijection
$(n,m)\rightarrow i$ such that we can label the basis states using a single
index $i$ such that
$\ket{\Phi_i} = \ket{\phi_{n_i}} \otimes \ket{\varphi_{m_i}}$.
In coordinate representation, we have
\begin{align}
 \label{eq:basisx}
 \braket{x}{\phi_{n_i}} =
\begin{cases}
 \sqrt{ \frac{2}{L_x} }\cos \left( \frac{n_i\pi x}{L_x} \right) \textrm{ if } n_i=1,3,5,...\\
 \sqrt{ \frac{2}{L_x} }\sin \left( \frac{n_i\pi x}{L_x} \right) \textrm{ if } n_i=2,4,6,...
\end{cases}
\end{align}
and
\begin{align}
 \label{eq:basisy}
 \braket{y}{\varphi_{m_i}} = \frac{\expo{-\frac{y^2}{2a_w^2}}}{\sqrt{2^{m_i}\sqrt\pi m_i! a_w}}H_{m_i}(y/a_w),\  m_i = 0,1,2,... \ ,
\end{align}
where $a_w = \sqrt{\hbar/ m\Omega_w}$ is a characteristic length of the system
and $H_{m_i}$ are Hermite polynomials.

After computing the matrix representation of $H_0$ in the chosen basis, we
diagonalize it and obtain it's eigenstates $\ket{\psi_i}$ and corresponding
energies $E_i$ which satisfy $H_0 \ket{\psi_i} = E_i \ket{\psi_i}$. Note that
$\ket{\psi_1}$ is the ground state, $\ket{\psi_2}$ the first excited state etc.
In the diagonalization process we also obtain a unitary transformation which
satisfies
\begin{align}
 U ( \ket{\phi_{n_i}}\otimes\ket{\varphi_{m_i}} ) = U \ket{\Phi_i} = \ket{\psi_i} \ .
\end{align}
Finally the wavefunctions of the lowest $N_{\textrm{ses}}$ single-electron
states $\psi_i(\mathbf r)$ are calculated and saved on a grid using 
\begin{align}
\psi_i(\mathbf r) = \braket{\mathbf r}{\psi_i} = \sum_{j=1}^{N_{xy}} U_{ij} \phi_{n_j}(x)\varphi_{m_j}(y) \ ,
\end{align}
where $N_{xy}$ is the number of basis states used for calculations. In actual
calculations we used approximately $120$ basis states in the $x$-direction and
$31$ in the y-direction so $n\in[1,120]$ and $m\in[0,30]$,
giving $N_{xy} = 120\times31 = 3720$. This is a large enough basis such that
numerical error due to the truncation is much smaller than the error due to
later truncation of many-body spaces. For this reason we will not investigate
convergence for the single-electron system in this paper.

\section{Many-electron Hamiltonian}
\label{sec:many}

We can write the many-electron Hamiltonian as a sum of two terms
$\mathcal H_e = \mathcal H_e^0 + \mathcal H_C$ where $\mathcal H_C$ only
contains the Coulomb interaction between electrons. Using the single-electron
eigenstates $\ket{\psi_i}\equiv \ket{i}$ as a basis, we can write the two terms
in second quantization as \cite{fetter2003quantum}
\begin{align}
 \mathcal H_e^0 & = \sum_{ij}\brainnf{i}{H_0}{j} d_i^\dagger d_j = \sum_{i} E_i d_i^\dagger d_i \\  \label{eq:VCoul}
 \mathcal H_C   & = \frac12 \sum_{ijrs} \brainnf{ij}{V_C}{rs} d_i^\dagger d_j^\dagger d_s d_r
\end{align}
where $d_i^\dagger$ ($d_i$) are fermionic creation (annihilation) operators of
an electron in state $\ket{i}$. The operators satisfy the usual fermionic
anti-commutation relation $\{ d_i,d_j^\dagger \} = \delta_{ij}$ and all other
anti-commutators are zero. The matrix element $\brainnf{ij}{V_C}{rs}$
in \eqref{eq:VCoul} is a double integral in the spacial variables and
involves integration with respect to the observation location $\mathbf r$
\begin{align}
 \label{eq:outerint}
 \langle ij| V |rs
 \rangle = \int d{\mathbf r} \; \psi_i^*({\mathbf r}) {\cal I}_{jr}({\mathbf r})
 \psi_s({\mathbf r})
\end{align}
and the integration with respect to the source location ${\mathbf r}^\prime$
\begin{align}
 \label{eq:innerint}
 {\cal I}_{jr}({\mathbf r}) = \int d{\mathbf r}^\prime \psi_j^*({\mathbf
 r}^\prime) V_C(\mathbf r, \mathbf r') \psi_r({\mathbf r}^\prime) \ ,
\end{align}
where $V_C$ is the Coulomb potential given by
\begin{align}
 \label{eq:CoulInt}
 V_C(\mathbf r,\mathbf r') = \frac{q^2/4\pi\epsilon}{\alg{\mathbf r -\mathbf r'}+\eta} \ ,
\end{align}
where $\eta$ is a small positive regularization parameter. The integrals in
\eqref{eq:outerint} and \eqref{eq:innerint} can not be done analytically due
to the nontrivial geometry so they are performed numerically using a Gaussian
quadrature scheme. We have to be careful with the numerical integration because
technically the wave functions reach infinity in the $y$-direction, although
exponentially decaying. We therefore have to find some sensible cutoff in the
$y$-direction where the amplitude of the eigenfunctions is close to zero. 
We used a grid size of $160\times120$ for the Gaussian integration. This grid
size is sufficiently large such that the numerical error in the Gaussian quadrature
is much smaller than the error due to basis truncations. We note however, that for
a larger magnetic field, a bigger grid might be required due to more rapid
fluctuations in the phase of the eigenfunctions $\psi_i(\mathbf r)$. 
To make sure that the $y$ cutoff is reasonable and the grid is sufficiently dense we checked the 
normalization of the eigenfunctions.

The Coulomb potential \eqref{eq:CoulInt} is integrable in the
origin, i.\ e.\ for ${\mathbf r}=\mathbf{ r'}$, in two dimensions, for
$\eta=0$. Therefore the integral \eqref{eq:innerint} is mathematically
convergent and the regularization parameter $\eta$ is theoretically not needed.  
However, due to the discretization of the two-dimensional space, working
in practice with $\eta=0$ can nevertheless cause problems in the numerical integration. A quick way around 
this problem is replacing ${\cal I}_{jr}({\mathbf r})$ with ${\cal \tilde I}_{jr}({\mathbf r})$ where
\begin{align}
 \nonumber
 \tilde{\mathcal I}_{jr}(\mathbf r)  \equiv 
  \int \slaufa{\psi_j^*(\mathbf r')-\psi_j^*(\mathbf r)}
  \frac{q^2/4\pi\epsilon}{\alg{\mathbf r -\mathbf r'}+\eta} \\  \label{eq:irj_best} 
 \slaufa{\psi_r(\mathbf r')-\psi_r(\mathbf r)} \dif \mathbf r' \ .
\end{align}
It's easy to show that the transformation ${\cal I}_{jr}({\mathbf r}) \rightarrow {\cal \tilde I}_{jr}({\mathbf r})$
leaves $\mathcal H_C$ unchanged and conveniently rids of us of the convergence problems we had with
$\mathcal I_{jr}(\mathbf r)$. The validity of this transformation does not depend on geometry or dimension
\cite[p.\ 63-65]{Jonasson2012_thesis}. We note that even though the limit $\eta\rightarrow0^+$ is well defined in
\eqref{eq:irj_best}, we still have to keep $\eta>0$ for numerical reasons. However, we can have $\eta$ much smaller
than if we used \eqref{eq:innerint} directly.

Now that we have the form of the many-electron Hamiltonian we need to find a suitable basis for the
many-electron Fock space. The natural choice is the occupation number basis $\slaufa{\ket\mu}$ where
\begin{align}
 \label{eq:fock}
 \ket{\mu} = \ket{n_1^\mu, n_2^\mu, n_3^\mu,\cdots ,n_\infty^\mu}\ ,
\end{align}
which means that $n_1^\mu$ particles are in state $\ket{\psi_1}$, $n_2^\mu$ in state $\ket{\psi_2}$ etc.
We use Latin indices for the single-electron states and Greek ones for the many electron states.
For fermions we have $n_i^\mu = 0$ or $n_i^\mu =1$. For example,
\begin{align} \label{bitstrings}
 \ket{0,1,1,0,1,0,0,...} = \ket{\psi_2}\otimes \ket{\psi_3} \otimes\ket{\psi_5} \ .
\end{align}
When doing calculations, the Fock space needs to be truncated by putting $\infty \rightarrow N\ntext{ses}$
in \eqref{eq:fock}, where $N\ntext{ses}$ is a finite positive integer. This means we are using a finite
number of single-electron states to construct the Fock space. This is the first truncation we perform
on Fock space. The corresponding number of many-electron states $N\ntext{mes}$ is $\binom{N\ntext{ses}}{N_e}$
where $N_e$ is the number of electrons. This rapid growth of $N\ntext{mes}$ obviously limits us to calculations
for a few electrons only.

To use this Fock basis we need some way to uniquely number the states. We need some mapping
$\Gamma: \ket\mu \rightarrow \mu$ where $\mu\in\mathbb{Z}^+$ and it's inverse $\Gamma^{-1}: \mu \rightarrow \ket\mu$.
There are many ways to construct $\Gamma$. The exact details will depend on factors such as whether or not all the states
$\ket\mu$ contain the same number of electrons. For a closed system the electron number is constant \cite{jonasson2012},
however an open system would have a varying number of electrons \cite{gudmundsson2012}. For this reason we will not go into
details of the form of $\Gamma$, but assume that we have such a mapping.  

We can now calculate the matrix representation of $\mathcal H_e$ in the $\slaufa{\ket\mu}$ basis using 
\begin{align}
 \nonumber
  \brainnf{\mu}{\mathcal H_e}{\nu} = & \delta_{\mu\nu} \sum_{i} n_i^\mu E_i 
 \\  \label{eq:HI_mat}
 + & \frac12 \sum_{ijrs} \brainnf{ij}{V_C}{rs}\brainnf{\mu}{d_i^\dagger d_j^\dagger d_s d_r}{\nu} \ ,
\end{align}
where $\brainnf{\mu}{d_i^\dagger d_j^\dagger d_s d_r}{\nu}$ is calculated using \cite{fetter2003quantum}
\begin{align}
  d_k \ket{\cdots n_k \cdots } = &
 \begin{cases}
  (-1)^{\gamma_k}\ket{\cdots 0 \cdots }, \ \ & \textrm{if } n_k = 1\\
  0, & \textrm{if } n_k = 0\\
 \end{cases} \ \\
 d_k^\dagger \ket{\cdots n_k \cdots } = &
 \begin{cases}
  0, & \textrm{ if } n_k = 1\\
  (-1)^{\gamma_k}\ket{\cdots 1 \cdots }, & \textrm{ if } n_k = 0\\
 \end{cases} \ ,
 \label{Fermi2}
\end{align}
with
\begin{align}
 \label{gamma}
 \gamma_k = \sum_{i=1}^{k-1} n_i \ .
\end{align}
The phase factor $(-1)^{\gamma_k}$ ensures that $d_i^\dagger$ and $d_i$ satisfy the fermionic anti-commutation relations.
Next we diagonalize $\mathcal H_e$ and find its eigenstates $\pket\mu$ and energies $\tilde E_\mu$.
In the diagonalization process we obtain a unitary transformation $\mathcal V$ which satisfies
\begin{align}
 \pket\mu = \sum_{\nu=1}^{N\ntext{mes}} \mathcal V_{\mu\nu} \ket\nu.
\end{align}
We distinguish between the many-body noninteracting and the many-body interacting states by
using an angular bracket for the kets of the first type, $|\mu\rangle$, and a rounded bracket for the 
kets of the second type, $|\mu )$, respectively.

This unitary transformation will be used extensively because it is much more efficient to perform
calculations in the $\slaufa{\ket\mu}$ basis and perform a unitary transformation to $\slaufa{\pket\mu}$,
rather than explicitly calculating and storing the many-electron eigenfunctions.
This means that every time we need $\pket\mu$ for calculations, we need to perform a unitary
transformation using a matrix that has the dimension $N\ntext{mes}\times N\ntext{mes}$.
This can be a problem since $N\ntext{mes}$ is a rapidly increasing function of $N_e$ and
$N\ntext{ses}$. For our calculations we use $N\ntext{ses}\simeq 50$ for two electrons,
resulting in $N\ntext{mes} = \binom{50}{2} = 1225$. For three electrons we use $N\ntext{ses}\simeq 30$,
resulting in $N\ntext{mes} = \binom{30}{3} = 4060$. The case for a single-electron is trivial since
$N\ntext{ses}=N\ntext{mes}$. For these values of $N\ntext{ses}$ and electron numbers we get a truncation
error that is smaller than the error due to the truncation of the electron-photon Fock space which is
covered in the next section. For this reason we will not go into discussion of convergence for the purely
electronic Fock space.

Before we go on and include interaction with a quantized EM field we note that if two Fock states
$\ket\mu$ and $\ket\nu$ do not have the same number of electrons, then
$\brainnf{\mu}{d_i^\dagger d_j^\dagger d_s d_r}{\nu}=\brainnf{\mu}{d_i^\dagger d_j}{\nu}=0$ for all $i$,$j$,$r$,$s$.
In other words the Coulomb interaction conserves the number of electrons.
This means that there exists a basis where $\mathcal H_e$ is block diagonal,
where each block consists of states with the same number of electrons. Therefore,
there exist unitary transformations $\mathcal V_{N_e}$ for each number of electrons
which has the same dimension as the block of $\mathcal H_e$ corresponding to $N_e$ electrons.
We can therefore use many small unitary transformations for each electron number instead of a
big one which works for all number of electrons. This can be a big boost in computation speed for large matrices.

\section{Inclusion of a quantized EM field}
\label{sec:EM}

Now suppose the system described in section \ref{sec:single} is subject to a single-mode quantized
electromagnetic field with vector potential $\mathbf A\ntext{EM}$. We can write the Hamiltonian as
\begin{align}
 \mathcal H = \mathcal H_e + \mathcal H\ntext{EM} + \mathcal H\ntext{int} \ ,
\end{align}
where $\mathcal H_e$ is the purely electronic Hamiltonian including the Coulomb interaction,
$\mathcal H\ntext{EM}$ is the free field photon term and $\mathcal H\ntext{int}$ contains the
electron-photon interaction. Ignoring the zero point energy, the free field term can be written as
$\mathcal H\ntext{EM} = \hbar\omega_p a^\dagger a$ where $\hbar \omega_p$ is the single photon energy
and $a$ ($a^\dagger$) is a bosonic annihilation (creation) operator. The electron-photon interaction
term can be split into two terms $\mathcal H\ntext{int} = \mathcal H\ntext{int}^{(1)} + \mathcal H\ntext{int}^{(2)}$
where
\begin{align}
 \label{eq:H1}
 \mathcal H\ntext{int}^{(1)} & \equiv \sum_{ij} \brainnf{\psi_i}{\frac{q}{2m} \svig{\boldsymbol \pi \cdot \mathbf A\ntext{EM} 
   + \mathbf A\ntext{EM}\cdot \boldsymbol \pi} }{\psi_j} d_i^\dagger d_j  \\ 
 \label{eq:H2}
 \mathcal H\ntext{int}^{(2)} & \equiv \sum_{ij} \brainnf{\psi_i}{\frac{q^2}{2m}  \alg{\mathbf A\ntext{EM}}^2}{\psi_j}d_i^\dagger d_j \ .
\end{align}
where $\boldsymbol \pi\equiv \mathbf p + q\mathbf A$ is the mechanical momentum. The term in \eqref{eq:H1}
is the paramagnetic interaction term and \eqref{eq:H2} is the diamagnetic term. To go further we need to
decide upon the form of $\mathbf A\ntext{EM}$. We assume that the single-mode photon wavelength is much
larger than characteristic length scales of the system. We can then approximate the vector potential
amplitude to be constant over the electronic system. Although related, this is not exactly the dipole
approximation since we will not omit the diamagnetic electron-photon interaction term. We can then
write the vector potential as
\begin{align}
 \label{eq:AEM}
 \mathbf A\ntext{EM} \simeq \mathbf{\hat e} A\ntext{EM}  (a+a^\dagger) = \mathbf{\hat e} \frac{\mathcal E_c}{q\Omega_wa_w}(a+a^\dagger) \ ,
\end{align}
where $\mathbf{\hat e}$ is a unit vector in the direction of the field polarization and
$\mathcal E_c \equiv qA\ntext{EM}\Omega_wa_w$ is the electron-photon coupling strength.

The strength of the photon-electron coupling is characterized by $A\ntext{EM}$,
the magnitude of which depends on the experimental setup. For a 3D Fabry Perot
cavity we would have $A\ntext{EM} = \sqrt{\hbar/(2\omega_pV\epsilon_0)}$ where
$V$ is the cavity volume. Another potential setup is a 1D transmission line
resonator \cite{Devoret2007} where it would be more appropriate to write $A\ntext{EM}$
in terms of the electric field vacuum fluctuation
$E\ntext{vac}^{\textrm{rms}} \equiv \sqrt{\brainnf{0}{\mathbf E\cdot \mathbf E}{0}}$ where
$\mathbf E  \equiv -\partial\mathbf A\ntext{EM}/\partial t$ and $\ket0$ is the lowest
eigenstates of $\mathcal H\ntext{EM}$. We would then have $A\ntext{EM}=E\ntext{vac}^{\textrm{rms}}/\omega_p$.

Using the approximation in Eq. \eqref{eq:AEM}, the expressions for $\mathcal H\ntext{int}^{(1,2)}$
in Eqs. \eqref{eq:H1}-\eqref{eq:H2} can be greatly simplified since we can pull $\mathbf A\ntext{EM}$
in front of the integrals and the commutator $[\mathbf A\ntext{EM},\boldsymbol \pi]$ is zero.
For the paramagnetic term, we get
\begin{align}
 \label{eq:Hint1}
 \mathcal H\ntext{int}^{(1)} \simeq 
 \mathcal E_c(a+a^\dagger)\sum_{ij} g_{ij} d_i^\dagger d_j \ .
\end{align}
where $g_{ij}$ is the dimensionless coupling between the electrons and the cavity mode defined by
\begin{align}\label{eq:gij}
 g_{ij} = \frac{a_w}{2\hbar} \mathbf{\hat e}\cdot \int d{\mathbf r} \left[ \psi_{i}^*({\mathbf
 r}) \left\{ \boldsymbol{\pi}\psi_{j}({\mathbf r}) \right\} \right.
 + \left. \left\{ \boldsymbol{\pi} \psi_{i}^*({\mathbf r})
 \right\} \psi_{j}({\mathbf r})
 \right] \ .
\end{align}
The dimensionless coupling $g_{ij}$ is closely related to the dipole transition moment
$\mathbf d_{ij}\equiv -q\brainnf{i}{\mathbf r}{j}$ according to
\begin{align}
 g_{ij} = i\svig{\frac{E_j-E_i}{\hbar\Omega_w}} \times \svig{\frac{\mathbf{\hat e}\cdot \mathbf d_{ij}}{q a_w}} \ .
\end{align}
A very accurate way to compute $g_{ij}$ is to calculate the integral in \eqref{eq:gij} analytically
in the original one electron basis $\slaufa{\ket{\phi_n}\otimes\ket{\varphi_m}}$ and perform a
unitary transformation into the $\slaufa{\ket{\psi_i}}$ basis. Another simpler method is to
store the $x$ and $y$ derivatives of $\psi_i(\mathbf r)$ on a grid and calculate \eqref{eq:gij}
using Gaussian quadrature. This method is less accurate but is easier to implement.

As for the diamagnetic term, we get
\begin{align}
 \mathcal H\ntext{int}^{(2)} \simeq \frac{\mathcal E_c^2}{\hbar\Omega_w} \horn{\svig{a^\dagger a + \frac12}+\frac12\svig{a^\dagger a^\dagger + aa}}\mathcal N^e \ ,
\end{align}
where $\mathcal N^e$ is the number operator in the electron Fock space. An interesting aspect
of $\mathcal H\ntext{int}^{(2)}$ is that it contains no dependence on the photon polarization
or geometry of the system.

A natural choice of basis for doing calculations is $\slaufa{\pket\mu\otimes\ket M}\equiv \slaufa{\ket{\breve \alpha}}$
where $\ket M$ are eigenstates of the photon number operator $a^\dagger a$, with $M$ the number of photons.
We will obviously need another bijection to label the states $\pket\mu\otimes\ket M$ with a single index $\alpha$.
The dependence of $\mu$ and $M$ on $\alpha$ is suppressed for easier reading. For the $\slaufa{\ket{\breve\alpha}}$
basis we use the lowest $N\ntext{mesT}\ll N\ntext{mes}$ Coulomb interacting eigenstates and photon states containing up
to $N\ntext{EM}$ photons, resulting in a total of $N\ntext{mesT}\times(N\ntext{EM}+1)$ states in the $\slaufa{\pket{\breve\alpha}}$
basis. This is the second time we truncate a many-body Fock space. Appropriate values of $N\ntext{mesT}$ and $N\ntext{EM}$ are
investigated in section \ref{sec:results}. 

Calculating matrix elements of $\mathcal H_e$ and $\mathcal H\ntext{EM}$ is straightforward in the $\slaufa{\ket{\breve \alpha}}$ basis. We get
\begin{align} \label{eq:Hemn}
 \brainnf{\mu;M}{\mathcal H_e}{\nu;N} & = \tilde E_\mu \delta_{\mu\nu}\delta_{MN} \\ \label{eq:HEMmn}
 \brainnf{\mu;M}{\mathcal H\ntext{EM}}{\nu;N} & = M \hbar\omega_p \delta_{\mu\nu}\delta_{MN} \ ,
\end{align}
where the shorthand $\ket{\mu;M} =\pket \mu \otimes \ket M$ has been used. For the paramagnetic interaction term we get
\begin{align}
 \nonumber
 \brainnf{\mu;M}{\mathcal H\ntext{int}^{(1)}}{\nu;N} = \mathcal E_c\sum_{ij} g_{ij} \pbrainnf{\mu}{d_i^\dagger d_j}{\nu}\brainnf{M}{a+a^\dagger}{N} \\
 \label{eq:H1mn}
= \mathcal E_c \mathcal G_{\mu\nu} \svig{\sqrt {M+1} \delta_{N,M+1} + \sqrt{N+1}\delta_{M,N+1}}\ ,
\end{align}
where we define
\begin{align}
 \mathcal G_{\mu\nu} \equiv \sum_{ij} g_{ij} \pbrainnf{\mu}{d_i^\dagger d_j}{\nu} = \sum_{ij} g_{ij} \brainnf{\mu}{\mathcal V^\dagger d_i^\dagger d_j\mathcal V}{\nu}\ ,
\end{align}
which is the many-electron generalization of $g_{ij}$. Its connection to the electron-photon coupling energy
in the Jaynes-Cummings model is explained in Ref.\ \cite{jonasson2012}. We will refer to $\mathcal G_{\mu\nu}$
as the dimensionless geometric coupling (DGC) between the electronic states $\pket\mu$ and $\pket\nu$.
As for the diamagnetic term we get
\begin{align}
 \nonumber
 \brainnf{\mu;M}{\mathcal H\ntext{int}^{(2)}}{\nu;N}
 = \frac{\mathcal E_c^2}{\hbar\Omega_w} N_\mu \delta_{\mu\nu} \left[ (N+\frac12)\delta_{MN}+\right. \\ \label{eq:H2mn}  \nonumber
  \left. \frac12 \sqrt{(M+1)(M+2)}\delta_{N,M+2} + \right. \\
\left. \frac12 \sqrt{(N+1)(N+2)}\delta_{M,N+2} \right],
\end{align}
where $N_\mu$ is the number of electrons in the state $\pket\mu$. The matrix elements of the total Hamiltonian
$\brainnf{\mu;M}{\mathcal H}{\nu;N}$ are obtained by adding \eqref{eq:Hemn}, \eqref{eq:HEMmn},
\eqref{eq:H1mn} and \eqref{eq:H2mn} together.

The final step is diagonalizing $\mathcal H$ and obtaining the allowed energies $\breve E_\alpha$
and the corresponding eigenstates $\pket{\breve \alpha}$ which are related to $\ket{\breve\alpha}$
by the unitary transformation
\begin{align}
 \pket{\breve \alpha} = \sum_{\beta} \mathcal W_{\alpha\beta}\ket{\breve\beta} \ ,
\end{align}
that is obtained in the diagonalization process. Again we use the right angular bracket for the 
basis states and the rounded bracket for the interacting states, this time the interaction being between
electrons and photons.

Expectation values of an observable $\mathcal A$ can then be calculated using
\begin{align}
 \left< \mathcal A \right> = \operatorname{Tr(\rho \mathcal A)} = \sum_{\alpha\beta} \brainnf{\breve\alpha}{\mathcal A}{\breve \beta} \hat \rho_{\beta\alpha}
 = \sum_{\alpha\beta} \pbrainnf{\breve\alpha}{\mathcal A}{\breve \beta} \breve \rho_{\beta\alpha}
\end{align}
where $\breve \rho$ ($\hat \rho$) is the density matrix of the system in the
$\pket{\breve \alpha}$ ($\ket{\breve \alpha}$) basis. The main advantage of working in the
$\ket{\breve \alpha}$ basis is that $\brainnf{\breve\alpha}{\mathcal A}{\breve \beta}$ is easy to calculate.
However it is very hard to truncate $\hat \rho$ effectively. Working in the $\pket{\breve \alpha}$ basis is
the exact opposite, $\pbrainnf{\breve\alpha}{\mathcal A}{\breve \beta} = \brainnf{\breve\alpha}{\mathcal W^\dagger \mathcal A \mathcal W}{\breve \beta}$
is expensive to calculate but it's easy to truncate $\breve \rho$ because if the system is in an energetically low state,
all of its biggest elements are concentrated in its top left corner (low $\alpha$ and $\beta$).
Note that, although $\pbrainnf{\breve\alpha}{\mathcal A}{\breve \beta}$ is relatively expensive to calculate,
it can be computed beforehand and saved.

Example of an interesting observable is the photon number operator $\mathcal N^{\textrm{ph}}=a^\dagger a$.
Its expectation value can be calculated using
\begin{align}
 \nonumber
 \left < \mathcal N^{\textrm{ph}}\right >  & = \sum_{\alpha\beta} \brainnf{\mu;M}{a^\dagger a}{\nu;N} \hat \rho_{\beta\alpha}\\
 & = \sum_{\alpha\beta} N_\alpha \delta_{MN} \delta_{\mu\nu} \hat \rho_{\beta\alpha} = \sum_{\alpha} N_\alpha \hat \rho_{\alpha\alpha} \ .
\end{align}
Another interesting observable is the charge density
\begin{align}
 \mathcal Q(\mathbf r) = -q \sum_{ij}\psi_i^*(\mathbf r) \psi_j(\mathbf r) d_i^\dagger d_j \ ,
\end{align}
the expectation value of which can be calculated using
\begin{align}
 \left< \mathcal Q \right>(\mathbf r) = \sum_{\alpha\beta} \sum_{ij} \psi_i^*(\mathbf r)\psi_j(\mathbf r) \pbrainnf{\mu}{d_i^\dagger d_j}{\nu} \delta_{MN} \hat \rho_{\alpha\beta} \ ,
\end{align}
where it is important to calculate $\pbrainnf{\mu}{d_i^\dagger d_j}{\nu}=\brainnf{\mu}{\mathcal V^\dagger d_i^\dagger d_j \mathcal V}{\nu}$
beforehand to avoid unnecessary repetitions.

\section{Results}
\label{sec:results}
For the results presented in this section we use $B = 0.1$~T, $\hbar\Omega_0=1.0$~meV, $L_x = 300$~nm, $m = 0.067m_e$
and $\epsilon = 12.4\epsilon_0$ (GaAs parameters). We choose $\omega_p$ such that the system is on resonance between
some chosen electron states $\pket\kappa$ and $\pket\lambda$ with detuning $\delta$, that is
$\hbar\omega_p = | \tilde E_\lambda-\tilde E_\kappa | + \delta$ where $\delta=0.01(\tilde E_\lambda- \tilde E_\kappa)$.
We refer to $\pket\kappa$ and $\pket\lambda$ as the active states. We use $\lambda>\kappa$, making $\delta$ positive so
that we are slightly over resonance. Choosing $\lambda<\kappa$ would give a negative $\delta$, resulting in a system
that is slightly under resonance. To distinguish between electron states with different number of electrons,
we use the notation $\pket{\mu}_{N_e}$ to denote the $\mu$-th electronic state containing $N_e$ electrons.
For example, $\pket{4}_2$ is the fourth lowest two electron state.

\begin{figure}[h!]
  \centering
  \includegraphics[width=0.44\textwidth]{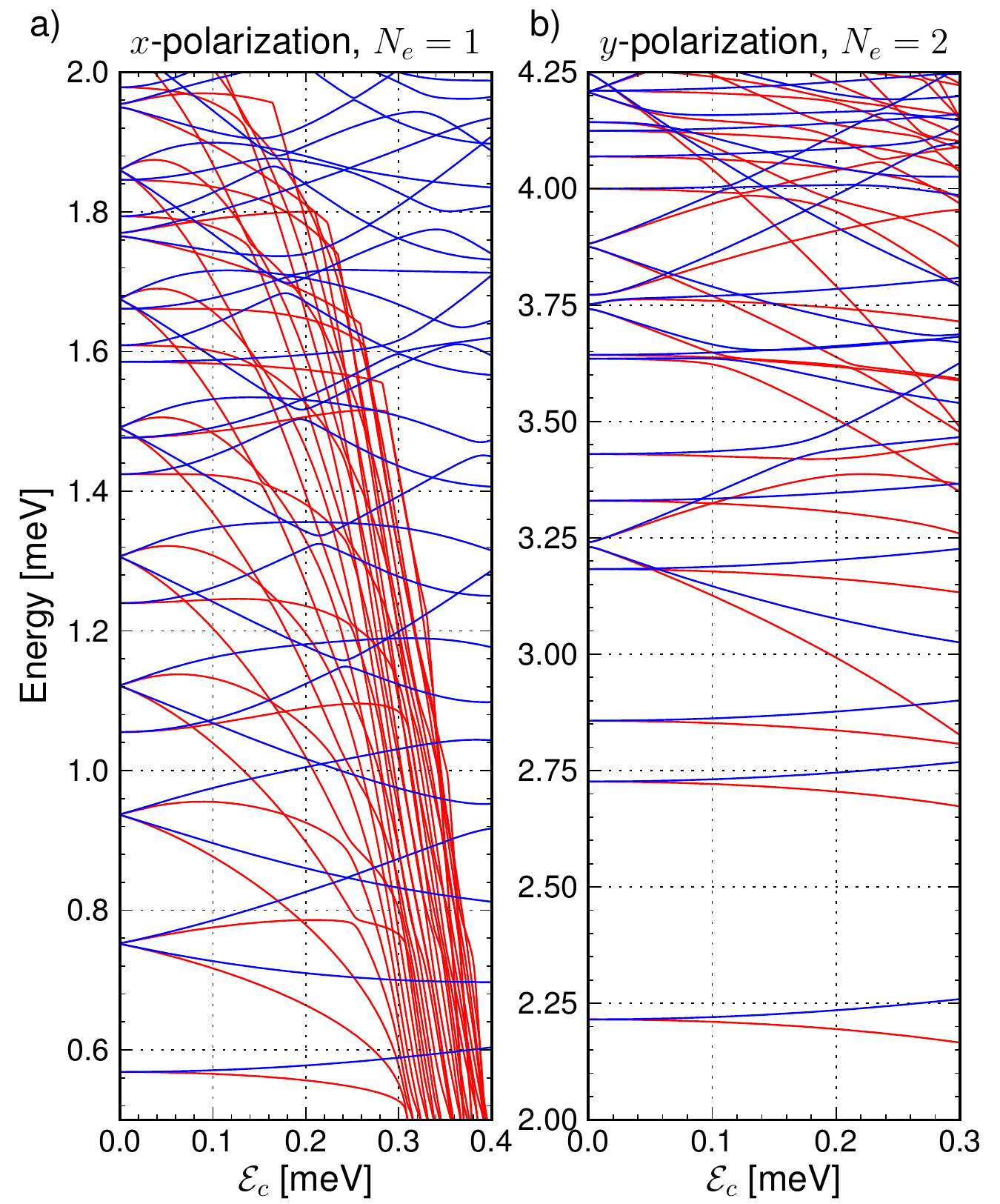}
  \caption{Energy spectra for the lowest $\sim 60$ states with one electron and $x$-polarization (a)
and two electrons and $y$-polarization (b). The diamagnetic $A^2$ term in the e-EM interaction
Hamiltonian is both included (blue) and omitted (red). In (a), the system is on resonance between
the one electron states $\pket1_1$ and $\pket2_1$ with a DGC strength of $\alg{\mathcal G_{12}}=0.290$
and $\hbar\omega_p=0.185$~meV. In (b), the system is on resonance between the two electron states
$\pket1_2$ and $\pket5_2$ with a DGC strength of $\alg{\mathcal G_{15}}=0.987$ and $\hbar\omega_p=1.025$~meV.
As can be seen from the figure, omitting the $A^2$ term does give accurate results for small $\mathcal E_c$,
while for large $\mathcal E_c$ the energy spectrum takes a steep dive downwards. This dive also takes place
in the two electron case, however it can't be seen in the chosen range of $\mathcal E_c$. There is no physical
significance in these dives since the results are highly divergent in those areas.}
  \label{fig:diamagnetic}
\end{figure}

Figure~\ref{fig:diamagnetic} shows the energy spectra of $\mathcal H$ as a function of the coupling strength
$\mathcal E_c$ for both $x$ and $y$ polarization. The importance of the diamagnetic interaction term is also
illustrated in Figure~\ref{fig:diamagnetic} by plotting the same energy spectrum, but omitting the diamagnetic term.
For small coupling, ignoring the diamagnetic term is a valid approximation. However, for higher coupling strength,
the model without the diamagnetic term starts exhibiting red shift with respect to the exact result. This red shift
becomes visible at around $\alg{\mathcal G_{\kappa\lambda}}\mathcal E_c/\hbar\omega_p \sim 0.1$, where the values of
$\lambda,\kappa$, $\alg{\mathcal G_{\kappa\lambda}}$ and $\hbar\omega_p$ are given in the figure text.
For even higher coupling strength, the results without the diamagnetic term start exhibiting an unphysical
downwards dive in energy. In this regime, the results are highly divergent with respect to $N\ntext{mesT}$.
However, keeping $N\ntext{mesT}$ constant, the results are convergent with respect to $N\ntext{EM}$.

To get an estimate of the numerical truncation errors we look at the 
relative variation of the energy of state $\pket{\breve\alpha}$ defined as
\begin{align}
 \label{eq:Rij}
 R^{(\alpha)}_{ij} \equiv \alg{\frac{E_i^{(\alpha)}-E_j^{(\alpha)}}{E_i^{(\alpha)}}}
\end{align}
where $E_i^{(\alpha)}$ is the energy of state $\pket{\breve\alpha}$ and $i$ refers to 
a specific parameter related to the size of the truncated Fock space. For example $i$ can be $N\ntext{mes}$, $N\ntext{mesT}$ or $N\ntext{EM}$. 
Typically, $j$ is the maximum value of that parameter which
can be used to obtain the numerical output in a
reasonable computing time. We vary $i$ and check the converge of the results. When changing the parameters
$i$ and $j$, all other accuracy parameters are kept constant. We also define the maximum error of the $N$ lowest states as
\begin{align}
 \label{eq:Rmax}
 R_{ij}\ttext{max} \equiv \max_{\beta\in[1,N]} R^{(\beta)}_{ij} \ .
\end{align}
The value we choose for $N$ depends on what we intend to use the states for, 
once we have obtained them. For calculating electron transport using the generalized master equation $64$ states
are typically used, so that is the value we will use for $N$  \cite{gudmundsson2012}. Our criteria for convergent
results is that the error is not visible on a graph such as in Figure~\ref{fig:diamagnetic}.
This condition translates into a maximum relative error of $\sim 10^{-3}$.

\begin{figure}[h!]
  \centering
  \includegraphics[width=0.47\textwidth]{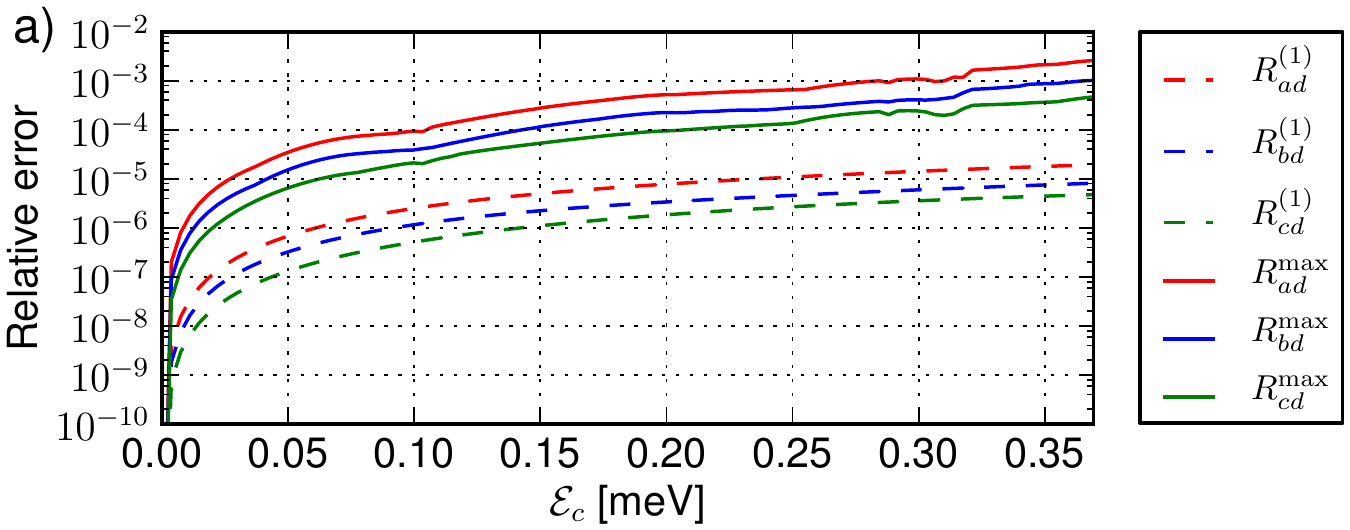}\\
  \includegraphics[width=0.47\textwidth]{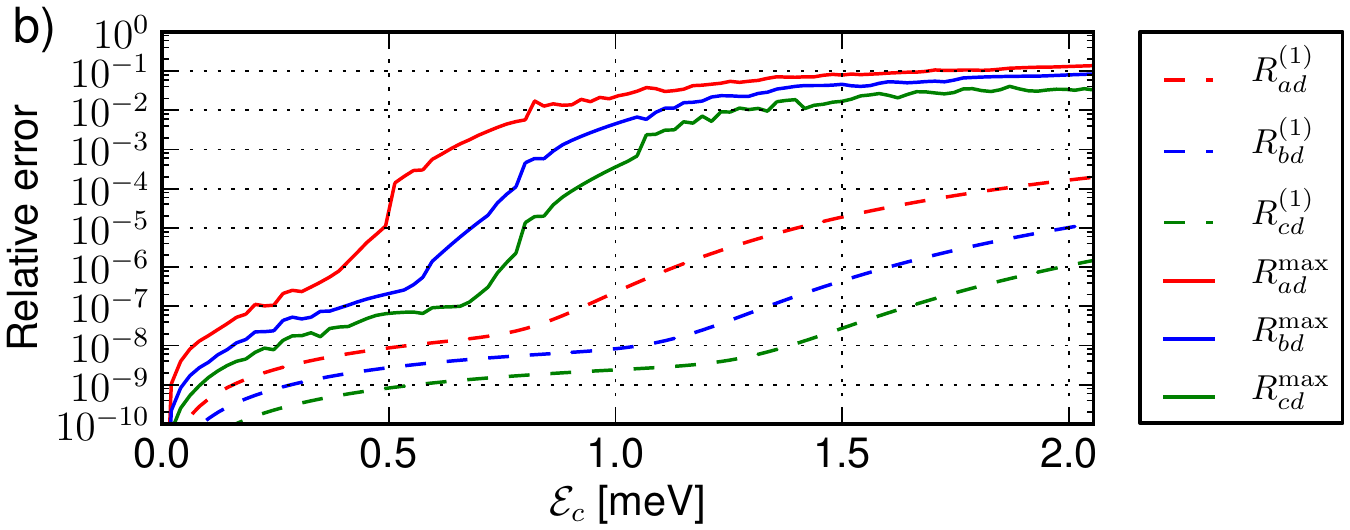}\\
  \caption{Convergence calculations with respect to $N\ntext{mesT}$ for $x$-polarization
           (a) and $y$-polarization (b). In (a),
           the system is on resonance between the one electron states $\pket1_1$ and $\pket2_1$ giving
           $\hbar \omega_p=0.185$~meV and $\alg{\mathcal G_{12}}=0.290$. The results are convergent up
           to $\mathcal E_c \simeq 0.37$ or $\alg{\mathcal G_{12}}\mathcal E_c/\hbar\omega_p\simeq 0.6$.
           In (b) the system is on resonance between the one electron states $\pket1_1$
           and $\pket5_1$ giving $\hbar \omega_p=1.03$~meV and $\alg{\mathcal G_{15}}=0.701$.
           The results are convergent up to $\mathcal E_c \simeq 1.05$ or
           $\alg{\mathcal G_{15}}\mathcal E_c/\hbar\omega_p\simeq 0.7$. For this run we have $a=100$, $b=150$,
           $c=200$ and $d=250$ (see equations \ref{eq:Rij} and \ref{eq:Rmax} for definition).
           The maximum number of photons is kept constant at $N\ntext{EM}=20$.}
  \label{fig:conv_1e_Nmes}
\end{figure}

Figure \ref{fig:conv_1e_Nmes} shows the relative error in the energy spectrum for one electron due to
the finite value of $N\ntext{mesT}$, that is the error due to the truncation of the electron part of
the joint electron-photon Fock space basis. From the figure we see that for $N\ntext{mesT}=200$,
results are convergent up to $\alg{\mathcal G_{12}}\mathcal E_c/\hbar\omega_p\simeq 0.6$ for $x$-polarization
and $\alg{\mathcal G_{15}}\mathcal E_c/\hbar\omega_p\simeq 0.7$ for $y$-polarization. From the figure we also
see that the error rises very rapidly for small $\mathcal E_c$ but as $\mathcal E_c$ becomes a considerable
fraction of $\hbar\omega_p$, the error increases much slower.

\begin{figure}[h!]
  \centering
  \includegraphics[width=0.47\textwidth]{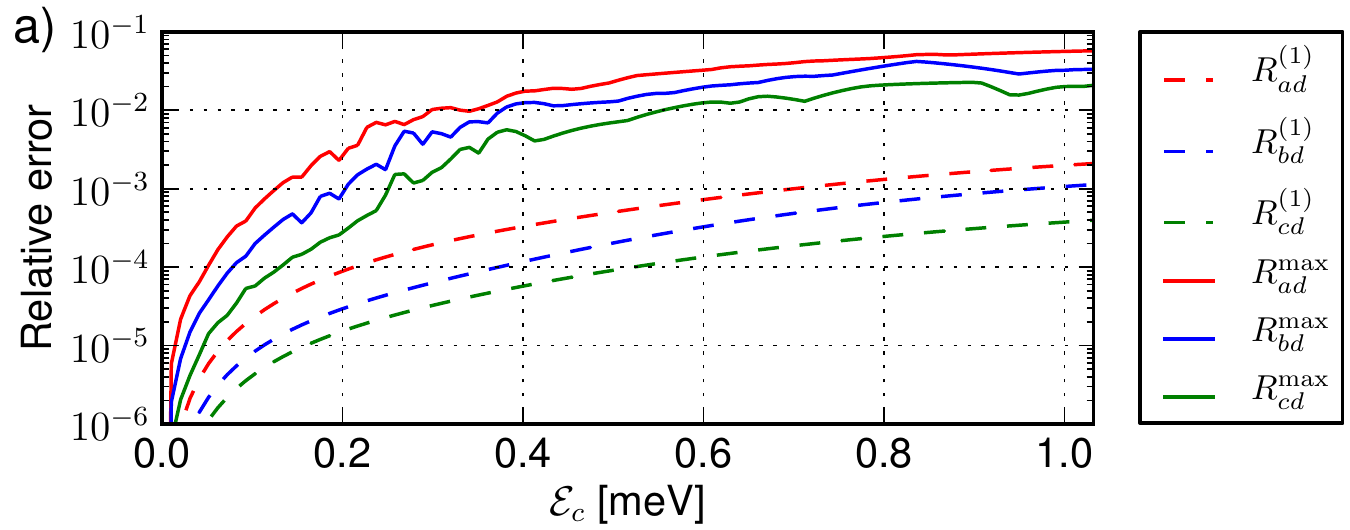}\\
  \includegraphics[width=0.47\textwidth]{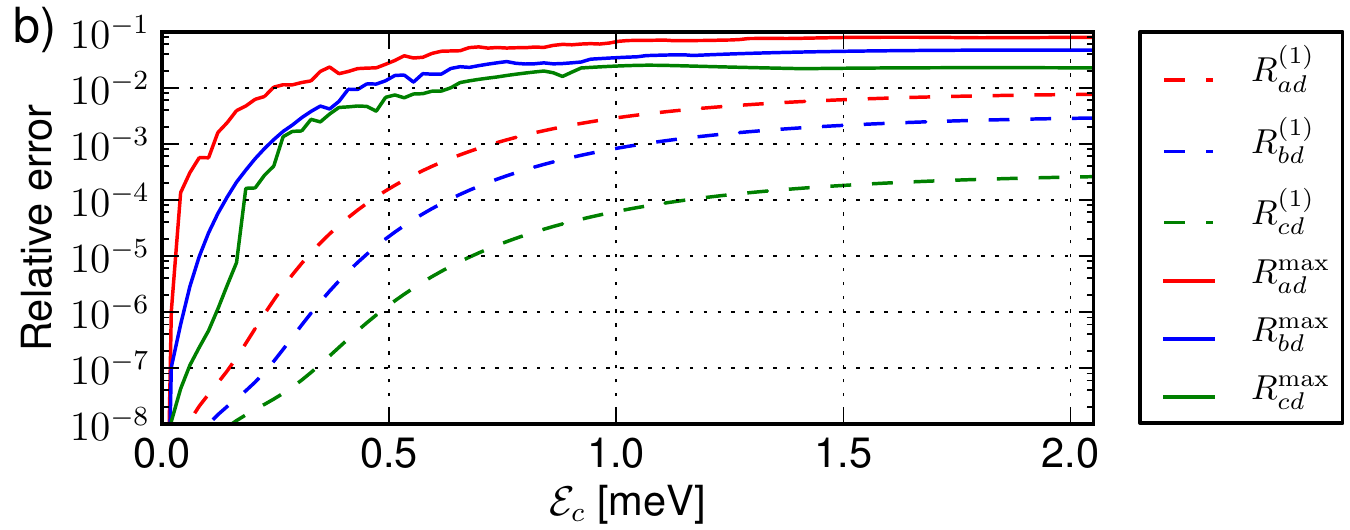}\\
  \caption{Convergence calculations with respect to $N\ntext{mesT}$ for $x$-polarization (a)
           and $y$-polarization (b). In (a), the system is on resonance
           between the two electron states $\pket1_2$ and $\pket2_2$ giving $\hbar \omega_p=0.516$~meV and
           $\alg{\mathcal G_{12}}=0.648$. The results are convergent up to $\mathcal E_c \simeq 0.25$ or
           $\alg{\mathcal G_{12}}\mathcal E_c/\hbar\omega_p\simeq 0.3$. In (b)
           the system is on resonance between the two electron states $\pket1_2$ and $\pket5_2$ giving
           $\hbar \omega_p=1.03$~meV and $\alg{\mathcal G_{15}}=0.987$. The results are convergent up to
           $\mathcal E_c \simeq 0.25$ or $\alg{\mathcal G_{15}}\mathcal E_c/\hbar\omega_p\simeq 0.24$.
           For this run we have $a=100$, $b=150$, $c=200$ and $d=250$ (see equations \ref{eq:Rij} and
           \ref{eq:Rmax} for definition). Other accuracy parameters are $N\ntext{ses}=50$ and $N\ntext{EM}=20$.}
  \label{fig:conv_2e_Nmes}
\end{figure}

Figure \ref{fig:conv_2e_Nmes} shows the relative error due to the finite value of $N\ntext{mesT}$ for two
electrons and both polarizations. For $N\ntext{mesT}=200$, the results are convergent up to
$\alg{\mathcal G_{12}}\mathcal E_c/\hbar\omega_p\simeq 0.3$ for $x$-polarization and
$\alg{\mathcal G_{15}}\mathcal E_c/\hbar\omega_p\simeq 0.3$ for $y$-polarization.
The same convergence calculations for $3$ electrons (not shown here) gives convergent results
for $\alg{\mathcal G_{13}}\mathcal E_c/\hbar\omega_p\simeq 0.25$ for $x$-polarization and
$\alg{\mathcal G_{15}}\mathcal E_c/\hbar\omega_p\simeq 0.25$ for $y$-polarization.

\begin{figure}[h!]
  \centering
  \includegraphics[width=0.48\textwidth]{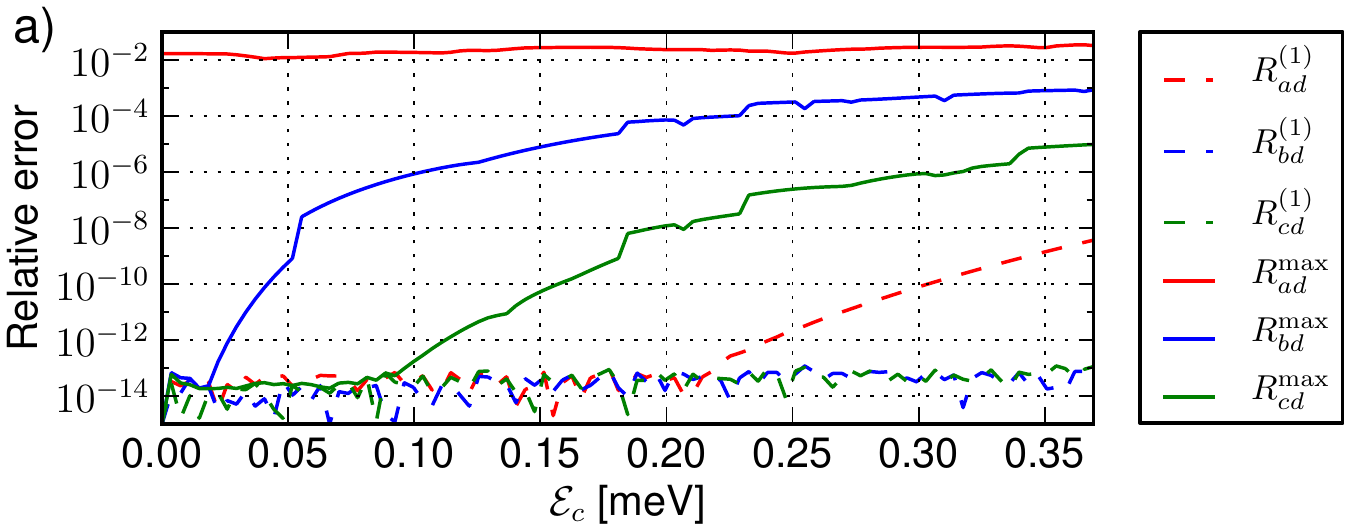}\\
  \includegraphics[width=0.48\textwidth]{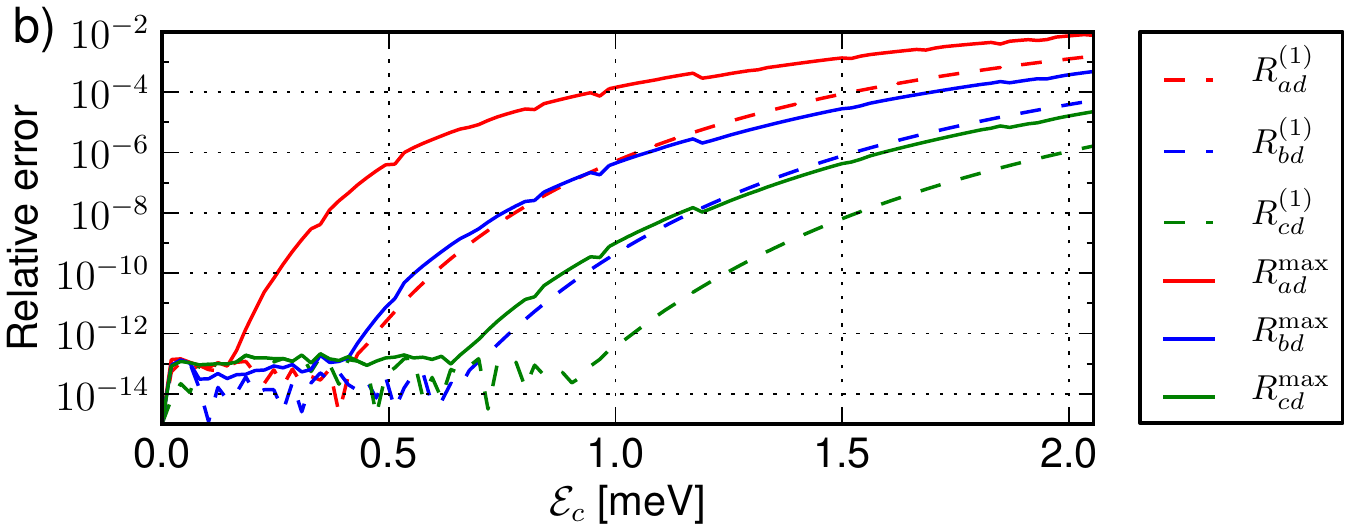}\\
  \caption{Convergence calculations with respect to $N\ntext{EM}$ for $x$-polarization (a)
           and $y$-polarization (b). Values of $\hbar\omega_p$ and $\alg{\mathcal G_{\lambda\kappa}}$
           are the same as in Figure~\ref{fig:conv_1e_Nmes} for both polarizations. We can see that for $N_{EM}=20$ (green),
           the results are acceptable for the whole range of $\mathcal E_c$ considered. For this run we have $a=10$, $b=15$,
           $c=20$ and $d=25$ (see equations \ref{eq:Rij} and \ref{eq:Rmax} for definition). The electron state number is kept
           constant at $N\ntext{mesT}=200$.}
  \label{fig:conv_1e_NEM}
\end{figure}

Figure~\ref{fig:conv_1e_NEM} shows the relative error due to the finite value of $N\ntext{EM}$,
that is the truncation of the photon part of the joint electron-photon Fock space basis. From
the figure we see that a modest value of $N\ntext{EM}=20$ is enough for the error to be $3-12$
orders of magnitude smaller than the $N\ntext{mesT}$ truncation error shown in figures \ref{fig:conv_1e_Nmes}
and \ref{fig:conv_2e_Nmes}. The results in Figure~\ref{fig:conv_1e_NEM} are for one electron but the two
and three electron cases (not shown here) exhibit the same behavior. The reason for this faster convergence
w.r.t $N\ntext{EM}$ is most likely that the electronic energy spectrum is much more dense, with a high amount
of energy crossings/anti-crossings, which requires a larger basis.

\begin{figure}[h!]
  \centering
  \includegraphics[width=0.48\textwidth]{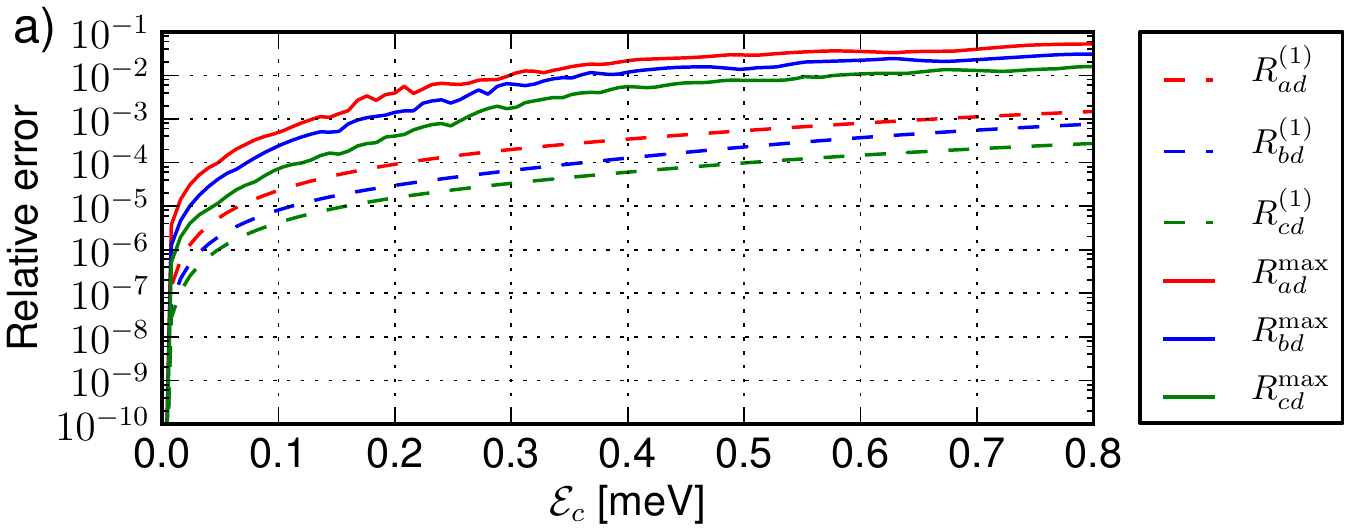}\\
  \includegraphics[width=0.48\textwidth]{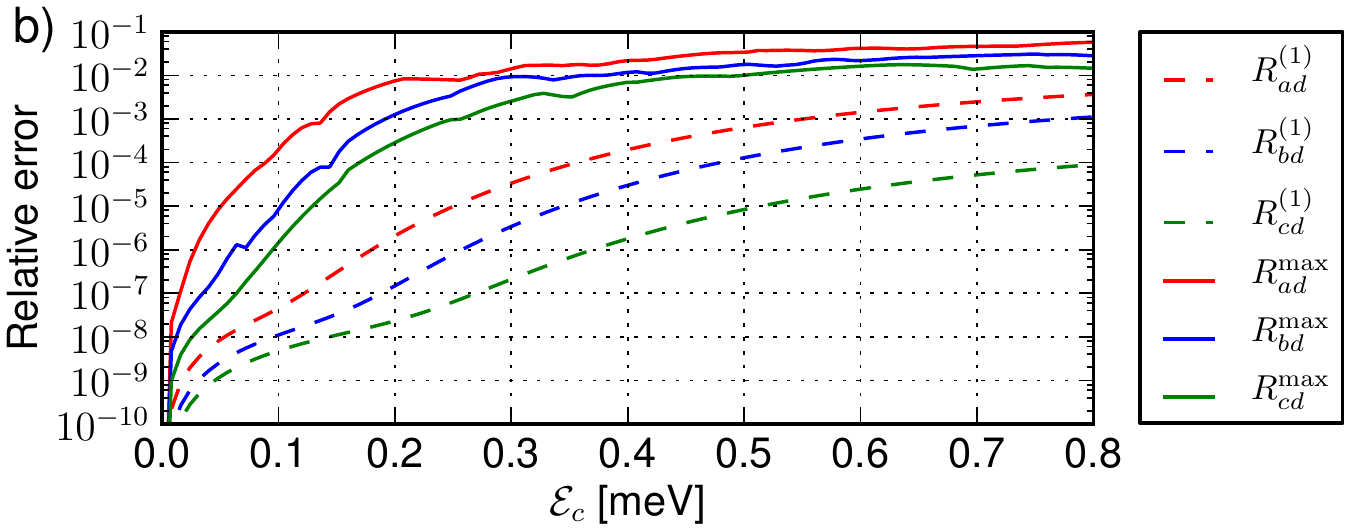}\\
  \caption{Convergence calculations for two electrons with respect to $N\ntext{mesT}$ for $x$-polarization
           (a) and $y$-polarization (b). For both polarizations,
           the system is off resonance with $\hbar \omega_p=0.4$~meV. In both cases, the results are convergent
           up to $\mathcal E_c\simeq 0.26$ or $\mathcal E_c/\hbar\omega_p\simeq 0.65$. For this run we have
           $a=100$, $b=150$, $c=200$ and $d=250$ (see equations \ref{eq:Rij} and \ref{eq:Rmax} for definition).
           Other accuracy parameters are $N\ntext{ses}=50$ and $N\ntext{EM}=20$.}
  \label{fig:conv_offres}
\end{figure}

Although in this paper we have put the photon frequency on resonance between two electronic states,
we are in no way forced to do so (see Ref.\ \cite{jonasson2012}). This motivates us to investigate
convergence for a system that is off resonance. Figure \ref{fig:conv_offres} shows convergence calculations
for a system that is off resonance and contains two electrons. From the figure we see that the results are
convergent up to $\mathcal E_c/\hbar\omega_p\simeq 0.65$ for both $x$ and $y$ polarizations. The reason
we use the ratio $\mathcal E_c/\hbar\omega_p$ rather than $\alg{\mathcal G_{\kappa\lambda}}\mathcal E_c/\hbar\omega_p$
is that when the system is off resonance, the concept of active states $\pket\lambda$ and $\pket\kappa$ has no meaning.

\section{Concluding remarks}
\label{sec:conclusions}

We have described a rigorous method to compute the many-body states of a multi level Coulomb interacting electronic system which 
also interacts with a single-mode quantized EM field. The model is exact in the sense that the only approximations are the finite
size of the single- and many-body bases and the finite size of grids on which single-electron eigenfunctions are stored.
The convergence with respect to these parameters is carefully controlled.

Due to the exact numerical nature of the model, calculations for arbitrarily strong photon-matter interaction
can in principle be performed with a big enough basis. Numerical results show that the main bottleneck is the
large number of electron states needed in the joint photon-electron many-body basis. Convergence with respect
to the number of photon states is much faster where $\sim 20$ states are sufficient to guarantee numerical error
that is $3-12$ orders of magnitude smaller than  the error caused by the electronic basis truncation with
$\sim200$ states. We have found that including the diamagnetic photon-electron interaction term drastically
improves convergence when the electron-photon coupling strength is considerable in size to the single photon
energy (ultrastrong coupling regime). Without the diamagnetic term, the model shows unphysical behavior
in the ultrastrong coupling regime due to divergent results.

\begin{acknowledgments}
The authors acknowledge financial support from the Icelandic Research and Instruments
Funds, the Research Fund of the University of Iceland, the National Science Council of
Taiwan under contract No. NSC100-2112-M-239-001-MY3. HSG acknowledges support from the
National Science Council in Taiwan under Grant No. 100-2112-M-002-003-MY3, 
from the National Taiwan University under Grants No. 10R80911 and 10R80911-2, and
from the focus group program of the National Center for Theoretical Sciences, Taiwan.
\end{acknowledgments}



%

%
%
\end{document}